\documentclass[a4paper,12pt]{article}
\usepackage{graphicx}
\global\arraycolsep=2pt 
\input{epsf}
\begin{document}
\makeatletter
\def\fmslash{\@ifnextchar[{\fmsl@sh}{\fmsl@sh[0mu]}}
\def\fmsl@sh[#1]#2{%
  \mathchoice
    {\@fmsl@sh\displaystyle{#1}{#2}}%
    {\@fmsl@sh\textstyle{#1}{#2}}%
    {\@fmsl@sh\scriptstyle{#1}{#2}}%
    {\@fmsl@sh\scriptscriptstyle{#1}{#2}}}
\def\@fmsl@sh#1#2#3{\m@th\ooalign{$\hfil#1\mkern#2/\hfil$\crcr$#1#3$}}
\makeatother
\thispagestyle{empty}
\begin{titlepage}

\begin{flushright}
hep-ph/0103012 \\
LMU 01/04 \\
\today
\end{flushright}

\vspace{0.3cm}
\boldmath
\begin{center}
{  \Large \bf The anomalous magnetic moment of the muon and radiative
  lepton decays}
\end{center}
\unboldmath
\vspace{0.8cm}
\begin{center}
  {\large Xavier Calmet, Harald Fritzsch and  Dirk Holtmannsp\"otter}
 \end{center}
 \vspace{.3cm}
\begin{center}
{\sl Ludwig-Maximilians-University Munich, Sektion Physik}\\
{\sl Theresienstra{\ss}e 37, D-80333 Munich, Germany}\\
\end{center}

\vspace{\fill}

\begin{abstract}
\noindent 
The leptons are viewed as composite objects, exhibiting anomalous
magnetic moments and anomalous flavor-changing transition moments. The
decay $\mu \to e \gamma$ is expected to occur with a branching ratio
of the same order as the present experimental limit.
\end{abstract}
to appear in Phys. Rev. D.
\end{titlepage}
Recently an indication was found that the anomalous magnetic moment of
the muon $\mu^+$ is slightly larger than expected within the standard
model \cite{Brown:2001mg}. The deviation is of the order of $10^{-9}$:
\begin{eqnarray}
  \Delta a_\mu &=& a_\mu(exp)-a_\mu(SM) = (4.3\pm 1.6) \times 10^{-9}.
\end{eqnarray}
For a review of the contribution of the standard model to the
anomalous magnetic moment of the muon see Ref.
\cite{Czarnecki:2001pv}. The observed effect (2.6 $\sigma$ excess)
does not necessarily imply a conflict with the standard model, in view
of the systematic uncertainties in the theoretical calculations due to
the hadronic corrections \cite{Yndurain}. If this result is confirmed
by further experimental data, it might be interpreted as the first
signal towards an internal structure of the leptons (see e.g. Ref.
\cite{Lane:2001ta}), although other interpretations (vertex corrections
due to new particles) are also possible \cite{allpapers}. A new
contribution to the magnetic moment of the muon can be described by
adding an effective term ${\cal L}^{eff}$ to the Lagrangian of the
standard model as follows:

\begin{eqnarray} \label{effL1}
  {\cal L}^{eff} &=& \frac{e}{2 \Lambda} \bar \mu
  \left(A + B \gamma_5 \right) \sigma_{\mu \nu}
  \mu F^{\mu \nu}, 
\end{eqnarray}
where $\mu$ is the muon field, $F^{\mu \nu}$ the electromagnetic field
strength, $\Lambda$ the compositeness scale and $A$ and $B$ are
constants of order one. We have included a $\gamma_5$-term in
view of a possible parity violation of the confining interaction.

The constants in ${\cal L}^{eff}$ depend on dynamical details of the
underlying composite structure. If the latter is analogous to QCD,
where such a term is induced by the hadronic dynamics, the constant is
of the order one, and the BNL result would give: $\Lambda \approx 2.5
\times 10^{7}$ GeV using

\begin{eqnarray} \label{amu1}
  \Delta a_\mu &=& \left(\frac{m_\mu}{\Lambda}\right),
\end{eqnarray}
assuming $|A|=1$. The $\gamma_5$-term does not contribute to the anomalous
magnetic moment.
  
The magnetic moment term (\ref{effL1}) has the same chiral structure
as the lepton mass term. Thus one expects that the same mechanism
which leads to the small lepton masses ($m_\mu \ll \Lambda$), e.g. a
chiral symmetry, leads to a corresponding suppression of the magnetic
moment. In this case the effective Lagrangian should be written as
follows:

\begin{eqnarray} \label{effL2}
  {\cal L}^{eff}&=& \frac{e}{2 \Lambda} \frac{m_\mu}{ \Lambda}
        \bar \mu \left(A + B \gamma_5 \right) \sigma_{\mu \nu}
  \mu F^{\mu \nu}, 
\end{eqnarray}
The contribution of the compositeness to the magnetic moment is in
this case given by
\begin{eqnarray} \label{amu2}
  \Delta a_\mu &=&  \left(\frac{m_\mu}{\Lambda}\right)^2.
\end{eqnarray}
Using the central value of $\Delta a_\mu$, one obtains:
$\Lambda\approx 1.61$ TeV, i.e.  $\Lambda$ is much smaller due to the
chiral symmetry argument \cite{Brodsky:1980zm}. The $95 \%$
confidence level range for $\Lambda$ is \cite{Lane:2001ta}
\begin{eqnarray} \label{rangeL}
 1.22\ \mbox{TeV} < \Lambda <  3.19 \ \mbox{TeV}.
\end{eqnarray}

If the leptons have a composite structure, the question arises whether
effects which are absent in the standard model, in particular
flavor-changing transitions, e.g. the decays $\mu \to e \gamma$ or
$\tau \to \mu \gamma$ arise.

In this note we shall study flavor changing magnetic-moment type
transitions which indeed lead to radiative decays of the charged
leptons on a level accessible to experiments in the near future.

We start by considering the limit $m_e=m_\mu=0$, i.e. only the third
lepton $\tau$ remains massive. Neutrino masses are not considered. In
this limit the mass matrix for the charged leptons has the structure
$m_{l^-}= m_\tau \mbox{diag} (0,0,1)$ and exhibits a ``democratic
symmetry'' \cite{Fritzsch:1999rd,Fritzsch:1994yx}. Furthermore there
exists a chiral symmetry $SU(2)_L \otimes SU(2)_R$ acting on the first
two lepton flavors. The magnetic moment term induced by compositeness,
being of a similar chiral nature as the mass term itself, must respect
this symmetry. We obtain
\begin{eqnarray} \label{effL3}
  {\cal L}^{eff} &=& \frac{e}{2 \Lambda} \frac{m_\tau}{ \Lambda}
        \bar \psi \tilde{M}\left(A + B \gamma_5 \right)  \sigma_{\mu \nu}
  \psi F^{\mu \nu}.
\end{eqnarray}
Here $\psi$ denotes the vector $(e, \mu, \tau)$ and $\tilde{M}$ is given by
$\tilde{M}=\mbox{diag}(0,0,1)$. 

Once the chiral symmetry is broken, the mass matrix receives non-zero
entries, and after diagonalization by suitable transformations in the
space of the lepton flavors it takes the form $M=\mbox{diag}(m_e,
m_\mu, m_\tau)$. If after symmetry breaking the mass matrix $M$ and
the magnetic moment matrix $\tilde{M}$ were identical, the same
diagonalization procedure which leads to a diagonalized mass matrix
would lead to a diagonalized magnetic moment matrix. However there is
no reason why $\tilde{M}$ and $M$ should be proportional to each other
after symmetry breaking. The matrix elements of the magnetic moment
operator depend on details of the internal structure in a different
way than the matrix elements of the mass density operator. Thus in
general the magnetic moment operator will not be diagonal, once the
mass matrix is diagonalized and vice versa. Thus there exist
flavor-non-diagonal terms (for a discussion of analogous effects for
the quarks see Ref.  \cite{Fritzsch:1999rd}), e.g. terms proportional
to $\bar e \ \sigma_{\mu \nu} \left ( A+ B \gamma_5 \right)\mu$. These
flavor-non-diagonal term must obey the constraints imposed by the
chiral symmetry, i.e. they must disappear once the masses of the light
leptons involved are turned off. For example, the $e-\mu$ transition
term must vanish for $m_e \to 0$. Furthermore the flavor changing
terms arise due to a mismatch between the mass density and the
magnetic moment operators due to the internal substructure. If the
substructure were turned off ($\Lambda \to \infty$), the effects should
not be present. The simplest Ansatz for the transition terms between
the leptons flavors $i$ and $j$ is $const. \sqrt{m_i m_j}/\Lambda$. It
obeys the constraints mentioned above: it vanishes once the mass of
one of the leptons is turned off, it is symmetric between $i$ and $j$
and it vanishes for $\Lambda \to \infty$. In this case the magnetic
moment operator has the general form:

\begin{eqnarray} \label{effL4}
  \! \! \! \!  \! \! \! \!
  {\cal L}^{eff} &=& \frac{e}{2 \Lambda} \frac{m_\tau}{ \Lambda}
        \bar \psi
\left( \begin{array}{ccc} \frac{m_e}{m_\tau}& C_{e \mu}
    \frac{\sqrt{m_e m_\mu}}{\Lambda} &
  C_{e \tau}
    \frac{\sqrt{m_e m_\tau}}{\Lambda}
    \\
  C_{e \mu}
    \frac{\sqrt{m_e m_\mu}}{\Lambda}
    &
\frac{m_\mu}{m_\tau}
    &
 C_{\mu \tau}
    \frac{\sqrt{m_\mu m_\tau}}{\Lambda}
    \\
C_{e \tau}
    \frac{\sqrt{m_e m_\tau}}{\Lambda}
    &
C_{\mu \tau}
    \frac{\sqrt{m_\mu m_\tau}}{\Lambda}
    &
1
    
\end{array} \right )
 \left(A + B \gamma_5 \right)  \sigma_{\mu \nu}
  \psi F^{\mu \nu}. \nonumber \\
\end{eqnarray}
Here $C_{ij}$ are constants of the order one. In general one may
introduce two different matrices (with different constants $C_{ij}$)
both for the 1-term and for the $\gamma_5$-term, but we shall limit
ourselves to the simpler structure given above.

Based on the flavor-changing transition terms given in eq.
(\ref{effL4}), we can calculate the branching ratios for the decays
$\mu \to e \gamma$, $\tau \to \mu \gamma$ and $\tau \to e \gamma$.
We find:
\begin{eqnarray}
  \Gamma(\mu \to e \gamma)&=& e^2  \frac{m_\mu}{4 \pi} \left(
    \frac{\sqrt{m_\mu m_e}}{\Lambda} \right)^2
    \left (\frac{m_\mu}{\Lambda}\right)^2
    \left (\frac{m_\tau}{\Lambda}\right)^2,
\\
  \Gamma(\tau \to \mu \gamma)&=& e^2 \frac{m_\tau}{4 \pi} \left(
    \frac{\sqrt{m_\tau m_\mu}}{\Lambda} \right)^2
    \left (\frac{m_\tau}{\Lambda}\right)^2
    \left (\frac{m_\tau}{\Lambda}\right)^2,
\\
  \Gamma(\tau \to e \gamma)&=& e^2 \frac{m_\tau}{4 \pi} \left(
    \frac{\sqrt{m_\tau m_e}}{\Lambda} \right)^2
    \left (\frac{m_\tau}{\Lambda}\right)^2
    \left (\frac{m_\tau}{\Lambda}\right)^2.
\end{eqnarray}

The corresponding branching ratios are, taking the constants equal to one:
\begin{eqnarray}
  \mbox{Br}(\mu \to e \gamma) &\approx& 2.8 \times 10^{-10},
  \\
   \mbox{Br}(\tau \to \mu \gamma) &\approx& 6.1 \times 10^{-10},
   \\
   \mbox{Br}(\tau \to e \gamma) &\approx& 3.0 \times 10^{-12},
 \end{eqnarray}
 using the central value of $\Delta a_\mu$ to evaluate $\Lambda$.
 One obtains the following ranges for the branching ratios
 \begin{eqnarray} \label{range}
 1.5\times 10^{-9}  &>\mbox{Br}(\mu \to e \gamma)&> 4.6 \times 10^{-12},
  \\
 3.3 \times 10^{-9} &> \mbox{Br}(\tau \to \mu \gamma) &> 1.0 \times 10^{-11},
   \\
 1.6\times 10^{-11}  &> \mbox{Br}(\tau \to e \gamma) &> 5.0 \times 10^{-14},
 \end{eqnarray}
 using the $95 \%$ confidence level range for $\Lambda$ (\ref{rangeL}).
  
 These ranges are based on the assumption that the constants of order
 one are fixed to one. The upper part of the range for the $\mu \to e
 \gamma$ decay given in (\ref{range}) is excluded by the present
 experimental limit: $\mbox{Br}(\mu \to e \gamma)<1.2 \times 10^{-11}$
 \cite{Groom:2000in}. Our estimates of the branching ratio should be
 viewed as order of magnitude estimates. In general we can say that
 the branching ratio for the $\mu \to e \gamma$ decay should lie
 between $10^{-13}$ and the present limit.
 
 The decay $\tau \to \mu \gamma$ processes at a level which cannot be
 observed, at least not in the foreseeable future. The decay $\tau \to
 e \gamma$ is, as expected, much suppressed compared to $\tau \to \mu
 \gamma$ decay and cannot be seen experimentally.
 
The experiment now under way at the PSI should be able to detect this
decay. If it is found, it would be an important milestone towards a
deeper understanding of the internal structure of the leptons and
quarks.

\section*{Acknowledgement}
  We should like to thank Z. Xing for useful discussions.

\end{document}